\documentclass[twocolumn,showpacs,preprintnumbers,amsmath,amssymb,aps,prl]{revtex4}
\pdfoutput=1
\usepackage{graphicx}
\usepackage{dcolumn}
\usepackage{bm}

\usepackage{verbatim}
\usepackage{bm} 
\usepackage{slashed}
\usepackage{mathdots}
\usepackage{color}%

\usepackage[colorlinks,urlcolor=blue,citecolor=blue]{hyperref}
\def\hhref#1{\href{http://arxiv.org/abs/#1}{arXiv:#1}} 

\allowdisplaybreaks[1]

\usepackage[all,cmtip]{xy}

\newcommand{\cpn}{${\mathbb C}{\mathbb P}^{N-1}$\,}
\newcommand{\cpone}{${\mathbb C}{\mathbb P}^{1}$}

\def\Z{\mathbb{Z}} 
\def\R{\mathbb{R}} 
 
\def\P{\mathbb{P}} 
\def\K{{\cal{K}}}

\def\be{\begin{equation}}
\def\ee{\end{equation}}

\def\Dslash{{\rlap{\raise 1pt \hbox{$\>/$}}D}}

\def\Z{\mathbb{Z}} 
\def\R{\mathbb{R}} 
 
\def\B{\mathbb{B}}

\def\Im{\text{Im}}
\def\Re{\text{Re}}
\def\tr{\text{tr}}

\def\hat{\widehat}
\def\bar{\overline}

\def\cB{{\mathcal B}}

\def\th{{\theta}}

\def\l{{\lambda}}

\def\s{{\sigma}}

\begin{document}
\title{Continuity and Resurgence:
towards a  continuum definition of the  \cpn model}
\author{Gerald V. Dunne}
\affiliation{Physics Department., University  of Connecticut, Storrs, CT, 06269}
\author{Mithat \"Unsal}
\affiliation{Department  of Physics and Astronomy, SFSU, San Francisco, CA 94132}

\begin{abstract}
 We introduce a non-perturbative continuum framework to study the dynamics of quantum field theory (QFT),  applied here to the $\mathbb C \P^{N-1}$ model,  using  \'Ecalle's  theory of resurgent trans-series, combined with the physical principle of continuity, in which
spatial compactification and  a Born-Oppenheimer approximation reduce QFT to quantum mechanics, while preventing all intervening rapid cross-overs or phase transitions. The reduced quantum mechanics contains the germ of all non-perturbative data, e.g., mass gap,  of the QFT, all of which are calculable. For \cpn, the results obtained at arbitrary $N$  are consistent with lattice and large-$N$ results. These theories are perturbatively non-Borel summable and possess the elusive IR-renormalon singularities. The trans-series expansion, in which perturbative and non-perturbative effects are intertwined, encapsulates the multi-length-scale nature of the theory, and eliminates all perturbative and non-perturbative ambiguities under consistent analytic continuation of the coupling. We demonstrate the cancellation of the leading non-perturbative ambiguity in  perturbation theory against the ambiguity in neutral bion amplitudes.
This provides a weak-coupling interpretation of the IR-renormalon, and a  theorem by Pham {\it et al} implies that the mass gap is a resurgent function,  for which resummation of the semi-classical expansion yields finite exact results.

\end{abstract}
\maketitle

The 1970s-80s witnessed an intensive research program on two-dimensional (2D) asymptotically free non-linear sigma models, motivated by their relevance to antiferromagnetic spin systems and  four dimensional (4D) QCD \cite{D'Adda:1978kp, Witten:1978bc,Jevicki:1979db,Affleck:1980mb, David:1980gi,Novikov:1984ac}.  Many results were found in the large-$N$ limit, particularly concerning  dynamical mass generation and chiral symmetry breaking. However, concrete results in finite-$N$ theories remain  scarce \cite{Bogli:2011aa}, and  even at large-$N$,  the microscopic mechanism by which a mass gap is generated remains open to date. The inheritance from this \'epoque  is a list of deep problems/puzzles  about \cpn.  A partial list includes:   
{\it i)} Invalidity of the dilute instanton gas approximation on $\R^2$. In a theory in which the instanton has size moduli, the dilute instanton gas is ill-defined, since it assumes that the typical inter-instanton separation is much larger then instanton size. This is a variant of Coleman's ``infrared embarrassment" problem. 
 {\it ii)} Perturbation theory leads to a non-Borel-summable divergent series even after regularization and renormalization. This is  a major reason why most  mathematicians would still consider QFT non-rigorous. Attempts to Borel resum perturbation theory yield a class of ambiguities associated with singularities in the complex Borel plane. While some such ambiguities  are cancelled by non-perturbative ambiguities 
 (associated with 2D instanton-anti-instanton events),  via a QFT version of the Bogomolny-Zinn-Justin mechanism \cite{Bogomolny:1980ur,ZinnJustin:1981dx}, there are other (more relevant) ambiguities associated with infrared (IR) renormalons  \cite{'tHooft:1977am, David:1980gi, Novikov:1984ac,Beneke:1998ui}, and there are no known (semi-classical or otherwise) 2D configurations with which these ambiguities may cancel. Therefore, (Borel resummed) perturbation theory by itself is ill-defined.
   {\it iii)} Precise connection between large-$N$ results and the instanton gas approximation \cite{Witten:1978bc,Jevicki:1979db,Affleck:1980mb,Aguado:2010ex}. 
 {\it iv)}  Microscopic mechanism underlying the large-$N$
mass gap for  $\mathbb C \P^{N-1}$:  $m_g = \Lambda=\mu e^{-S_I/N}  =  \mu e^{-\frac{4\pi}{g^2 N}}$, where  $\mu$ and  $\Lambda$ are  the renormalization and strong scale. 

 In this Letter we revisit these problems using new physical and mathematical analytical tools. Our strategy is to use spatial  (non-thermal) compactification and continuity to render the theory calculable,  then apply \'Ecalle's resurgence formalism \cite{Ecalle:1981} to unify perturbative and non-perturbative aspects in a manner that respects analytic continuation in the coupling, yielding exact unambiguous answers for physical observables.

The \cpn model is described by a quantum  field $n(x)$ in the coset space 
$  \frac{U(N)}{U(N-1) \times U(1)} $. The action is   
\begin{eqnarray}
 S=\int d^2 x \left[ \frac{2}{g^2}   \left(D_\mu n\right)^\dagger D_\mu n  
 -\frac{i  \Theta }{2\pi} \, \epsilon_{\mu\nu} (D_\mu n)^\dagger D_\nu n 
 \right]
\label{action}
\end{eqnarray}
where $D_{\mu}= \partial_{\mu} +i A_{\mu}$,  $A_{\mu}$ is an auxiliary field, and $ \Theta$ is the topological angle.  \cpn has a global $U(N)$ symmetry, and a local $U(1)$ gauge redundancy, $n \to e^{i \alpha(x)}n$.  
We parametrize \cpn by locally splitting the $n$-field  into a phase and modulus,  $n_i = e^{i \varphi_i} r_i$, involving $(N-1)$ phase fields $\{\varphi_1, \ldots,  \varphi_N\}, \; \sum_{i=1}^N \varphi_i=0$ mod($2\pi$) by $U(1)$ gauge redundancy,  and $(N-1)$ modulus fields  $\{ r_1, \ldots, r_N \},  \; \sum_{i=1}^N r_i^2=1$.  We also  consider adding  $N_f$ species of Dirac fermions, although here we  mostly deal with the bosonic $N_f=0$ theory.  

Each  $-\partial_\mu \varphi_i  \equiv {\cal A}_{\mu,i} $ transforms as a ``gauge" connection, 
${\cal A}_{\mu,i}  \longrightarrow    {\cal A}_{\mu,i}  -   \partial_\mu  \alpha$, 
and we refer to it as the ``$\sigma$-connection".   With spatial compactification on $\R \times \mathbb S^1_L$,   we define a new order parameter, the  ``$\s$-connection holonomy'',  making a circuit around the compact direction: 
 \begin{eqnarray}
&& (^L\Omega)_j (x_1) : = e^{ i \int_{0}^{L}  dx_2 \;  {\cal 
A}_{2,j} } = e^{  i (\varphi_j (x_1, 0) -\varphi_j(x_1 , L)) }
\cr 
&&^L\Omega(x_1) : =   {\rm Diag}\Big[  (^L\Omega)_1 (x_1) , \dots,  (^L\Omega)_N(x_1)  \Big]
\label{hol}
\end{eqnarray} 
Under  an aperiodic global  gauge rotation, the line operator 
  $^L\Omega(x_1)  \to  \;  e^{i \frac{2 \pi k}{N}} 
\,  ^L\Omega(x_1),  \quad k=1, \ldots, N $, 
reflecting the theory's global $\Z_N$ center-symmetry.   (The fact that it is not $U(1)$  follows 
from the constraint, $\sum_{i=1}^N \varphi_i=0$ mod($2\pi$), i.e.,  $\det ^L\Omega(x_1) =1$.)
The  matrix-valued gauge invariant  $\s$-holonomy,  $^L\Omega(x_1)$, plays an analogous role to the Wilson line 
in non-abelian $SU(N)$ gauge theory. Crucially,  this operator contains more refined  data than  the familiar $U(1)$ Wilson line in the \cpn model: $W= e^{i   \int_{0}^{L}  dx_2  A_{2} }$.  
 
 A typical  classical background of the $\s$-connection holonomy  is 
 $^L\Omega_{\rm b}=  {\rm Diag} \Big[e^{2\pi i\mu_1}, e^{2\pi i\mu_2}, \ldots, e^{2\pi i\mu_N} \Big]$. Classically, this background is equivalent to imposing  twisted boundary conditions for the \cpn fields, of the form $n(x_1, x_2+L) = \;  ^L\Omega_{\rm b}n(x_1, x_2)$ \cite{Bruckmann:2007zh, Brendel:2009mp}.  Undoing the twist by a field redefinition is equivalent to the substitution,  
$  \partial_\mu  \rightarrow \partial_\mu   + i \; \delta_{\mu 2} \;   \frac{2\pi}{L} \;  {\rm Diag} \Big[\mu_1, \mu_2, \ldots, \mu_N \Big]$, analogous to turning on a Wilson line   in compactified  $SU(N)$  gauge theory (and further reason to call  $ - \partial_\mu \varphi_i = {\cal A}_{\mu,i}$  $\s$-connection).  In the small-$L$  weak coupling regime, the quantum mechanical stability of a given background can be  determined via a one-loop analysis of the potential for  the holonomy (\ref{hol}), similar to gauge theory.  
 Integrating out weakly coupled  Kaluza-Klein   modes,  we find
    \begin{eqnarray}
    &&V_{\pm} [ {^L}\Omega]  =   \frac{2}{ \pi L^2}  \sum_{n=1}^{\infty} \frac{1}{n^2}   \left( -1 + N_f (\pm1)^n \right)
 (|\tr \, {^L}\Omega^n |-1)  \nonumber
 \label{spotential} 
\end{eqnarray} 
where $-$ (+) refers to thermal  (spatial) compactification, where fermions have anti-periodic (periodic)  boundary conditions. 
For $N_f \geq 0$,  the {\it $N$-fold degenerate} minima in the thermal case   are  $ {^L} \Omega_0^{\rm thermal} = e^{i \frac{2\pi k}{N} }{\bf 1}_N, k=1,\ldots N$, a clumped configuration of holonomy eigenvalues. 
In sharp contradistinction, for  $N_f \geq 1$ and 
in the spatial case, the minimum  is  unique,  $ {^L} \Omega_0^{\rm spatial}  =    {\rm Diag} \left( 
1,  e^{i \frac{2\pi}{N} }, \ldots,  e^{ i \frac{ 2\pi (N-1)}{N}} \right)$, a non-degenerate, $\Z_N$-symmetric holonomy, similar to QCD(adj) \cite{Unsal:2007vu, 
Unsal:2007jx}. The $N_f=1$ case follows from non-perturbative effects \cite{Dunne:2012ae}.

Since there are no phase transitions  (for finite-$N$) on $\R \times \mathbb S^1_L$,  one may wonder what qualitative differences these two different backgrounds entail. 
In the thermal case, the potential at the minimum is the free energy density  of the hot \cpn model, 
${\cal F} = V_{-} [^L\Omega_0^{\rm thermal}] = -(2N-2)  \frac{\pi}{6} \left( 1+ \frac{N_f}{2}\right) T^2    \sim O(N^1)$,  the  Stefan-Boltzmann result; whereas in the cold regime, ${\cal F}  \sim O(N^0)$, because the spectral density of physical states is $O(N^0)$.   There is a rapid cross-over from a hot deconfined regime to the cold confined regime at the strong scale at finite-$N$, which becomes a sharp phase transition at $N=\infty$. With spatial compactification, the ``free energy" density at small-$L$ is ${\cal F}  \sim O(N^0)$, just like the cold regime of the thermal theory.  Therefore, there is no intervening rapid crossover (finite-$N$)  or  phase transition $(N=\infty)$ as one dials the radius from large to small.  This is the reason that the spatially  compactified theory  provides, in the small $L$ regime, a  weak-coupling semi-classical window into the  confined  regime \cite{Unsal:2007vu, Unsal:2007jx}, whereas the thermally compactified theory  provides, at small $ \beta$, a  weak coupling semi-classical window of the deconfined regime \cite{Affleck:1980mb}.  

 In the pure bosonic theory  ($N_f=0$) in which there is no distinction between 
 the thermal and spatial compactification,  a $\Z_N$-symmetric background is unstable. 
 However, one can define a deformed bosonic \cpn (analogous to deformed YM) by introducing
 heavy fermions $m \gg \Lambda$, 
 so that the theory at distances larger than $m^{-1}$ emulates the pure  bosonic theory. 
 Then, render the KK-modes sufficiently high such that  the heavy fermions appear  light with respect to the  KK-modes,   i.e.,  $m \lesssim  \frac{2 \pi}{LN}$. Thus, from the point of view of the one-loop potential, we can use the result of the massless theory,   and at the same time, at distances larger than $m^{-1}$, the theory is the bosonic  \cpn model on    a stable
 $\Z_N$-symmetric background.  
 
 In \cpone,   in the small-$L$ regime, the Lagrangian associated with the zero mode of the KK-tower is ($\xi\equiv\frac{2\pi}{N\, L}$):
 \begin{equation}
S^{\rm  zero}=   \frac{L}{2 g^2} \int_{\R}  (\partial_t \theta)^2 + 
\sin^2 \theta   (\partial_t \phi)^2  + \xi^2   \sin^2 \theta, \;  
\label{zeroth}
\end{equation} 
where the kinetic term  describes a particle on $S^2 \sim$ \cpone,  and the potential follows from the  $\Z_2$ stable background. 
This action  has a semi-classical kink-instanton solution,  which we call  $\K_1$, interpolating from $\theta=0$ to $\theta =\pi$, and with  action  $ S_1=   \frac{L}{g^2}  (2  \xi)=   
 \frac{4\pi}{g^2} \times (\mu_2-\mu_1)  = \frac{S_I}{2}$. Its topological charge is  $Q= \frac{1}{2}$. 
 There is also an independent kink-instanton, $\K_2$,  with  action $S_2=  \frac{S_I}{2}$, interpolating from  $\theta=\pi$ to $\theta =0$, also with topological charge  $Q= \frac{1}{2}$. 
It is important to note that $\K_2$ is not the anti-kink $\bar  \K_1$, which has  $Q=- \frac{1}{2}$. 
The kink-instanton $\K_2$ is  associated  with the affine root of the $SU(2)$ algebra \cite{Bruckmann:2007zh, Brendel:2009mp}. 

This construction generalizes to \cpn: there are $N$ types of kinks, associated with the extended root system of the $SU(N)$ algebra,  each of which carries a topological charge $Q= \frac{1}{N}$. The amplitude of $\K_i$ has a non-perturbative factor
\begin{equation}
\K_i: e^{-S_{0}}=  e^{- \frac{4\pi}{g^2N} }  = e^{- \frac{4\pi}{\lambda} }   \sim  e^{-\frac{S_I}{N}},  \;\;  i=1, \dots, N
\label{ki}
\end{equation}
It is crucial to note the appearance of the 't Hooft coupling, $g^2 N \equiv \lambda$, in the amplitude. Thus, the  kink-instantons  are exponentially more relevant that the 2D instanton:  $e^{- \frac{4\pi}{g^2N} } \gg  
e^{- \frac{4\pi}{g^2} } $.

At second order in the semiclassical expansion,  there are self-dual and non-self-dual configurations. The non-self-dual topological molecules are in one-to-one correspondence with the non-vanishing elements of the extended Cartan matrix $\hat A_{ij}$ of $SU(N)$. For each entry $\hat A_{ij}<0$  of the extended Cartan matrix, there exists a \emph{charged  bion}, $\cB_{ij} \sim [\K_i \bar \K_j]$, which plays a crucial role in the mass gap of the theory with $N_f \geq 1$, similar to  the magnetic bion in QCD(adj) on $\R^3 \times S^1$ \cite{Unsal:2007jx}.
\vspace{-4mm}
\begin{figure}[htb] \centering{\includegraphics[scale=0.2]{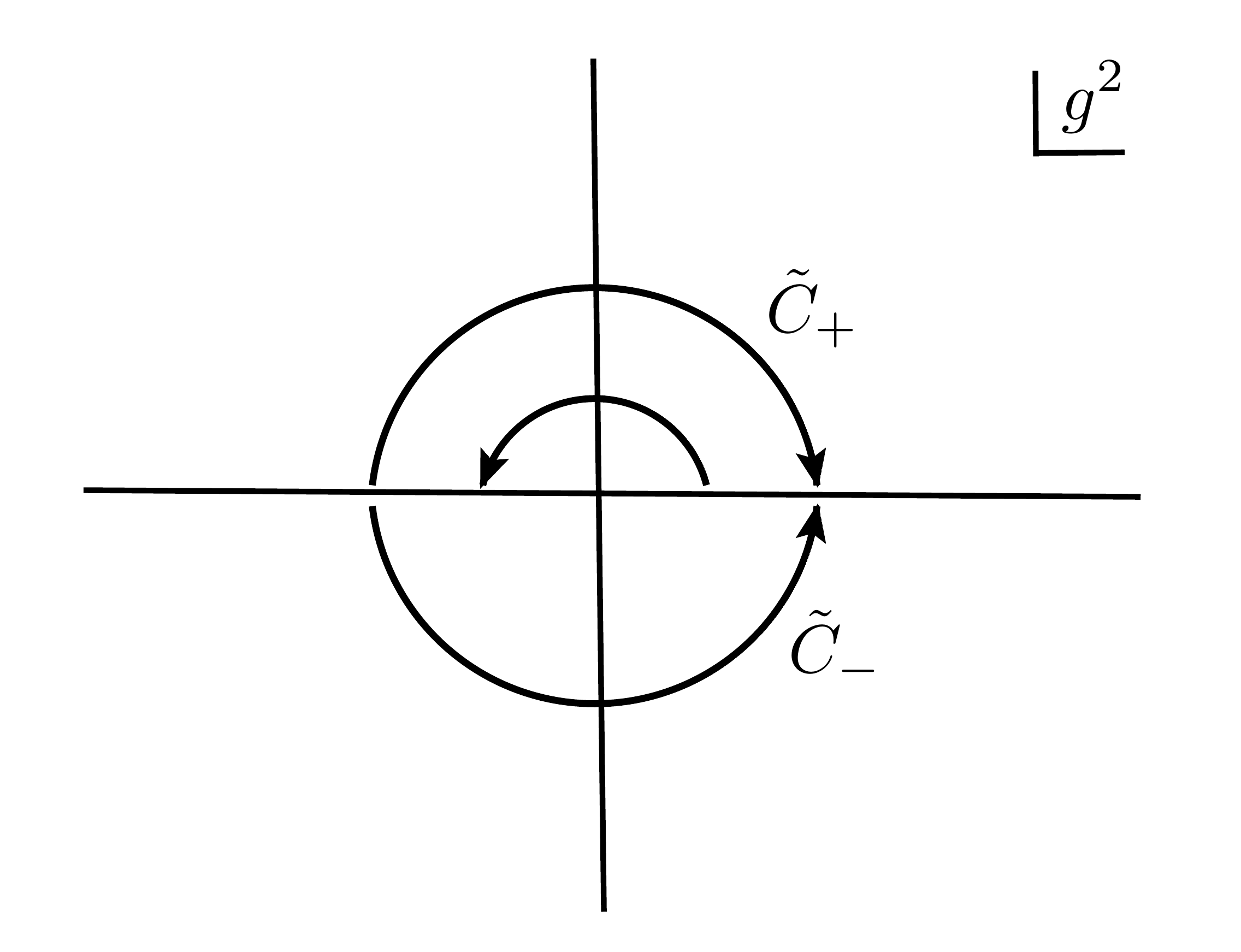}} 
\vspace{-4mm}
\caption{Defining left (right)  Borel sum  $\B_{\th=0^{\pm}}$, and left (right) neutral bion amplitude $ [\cB_{ii}]_{\th=0^{\pm}}$. The $g^2 >0$ line is a Stokes ray, the mathematical reason underlying the divergence of perturbation theory. } \label{fig:continuation2} \end{figure}
\vspace{-4mm}
For each diagonal entry, $\hat A_{ii}>0$, there exists a \emph{neutral bion}, $\cB_{ii} \sim  [\K_i \bar \K_i]$, with zero topological charge, and indistinguishable from the perturbative vacuum 
\cite{Argyres:2012ka, Argyres:2012vv}.
  The $\cB_{ii}$ generate a repulsion among the eigenvalues of the  holonomy (\ref{hol}).    For $g^2 >0$,   the constituents of the neutral bion  interact attractively at short distances and the quasi-zero mode integral yields an amplitude which is naively meaningless.  However, 
  this is actually a reflection of the  ``Stokes phenomenon''. 
We can evaluate, in the $N_f=0$ theory, the neutral bion amplitude as shown in Fig.~\ref{fig:continuation2}. 
First, take $g^2 \rightarrow -g^2$, where the neutral bion amplitude is well-defined. Then, 
analytic continuation  along $\tilde C_{\pm}$  to $\theta=0^{\pm}$ yields  {\it left} and {\it right} amplitudes 
    \begin{align} 
  [\K_{i}  \bar \K_{i} ]_{\th=0^{\pm}} =& \Re\,[\K_{i}  \bar \K_{i} ] + i \,
   \Im\, [\K_{i}  \bar \K_{i} ]_{\th=0^{\pm}}  \cr
   = &  \left(  \log   \left(   \frac{\lambda}{8 \pi } \right)  - \gamma \right) 
\frac{16 }{\lambda}
   e^{-\frac{8 \pi}{\l}} 
 \pm  i  \frac{16 \pi}{\lambda}   e^{-\frac{8 \pi}{\l}} 
    \qquad 
    \label{bion-ambiguity}
     \end{align}
The  {\it absence} of a well-defined  $\theta\to0$ limit means that the semi-classical expansion by itself is also ill-defined. \'Ecalle's resurgent approach is to simultaneously apply this analytic continuation to the Borel summation of the perturbative sector {\it and} to the non-perturbative sector, in such a way that all ambiguities cancel, yielding an unambiguous and exact ``trans-series'' result.
This mathematical technique has only been partially explored in QFT \cite{Marino:2012zq,Argyres:2012ka}, as most semi-classical studies only capture the first order.  In quantum mechanics, this effect is studied  in  \cite{Bogomolny:1980ur, ZinnJustin:1981dx}. 
Surprisingly, in QFT there appears to be universal behavior for the jump in the amplitude of neutral topological defects, arising from analytic continuation of the quasi-zero-mode integrals \cite{Argyres:2012ka, Argyres:2012vv, Unsal:2012zj,Dunne:2012ae}.  

The  low energy Hamiltonian for  (\ref{zeroth}), dropping  $\phi$-angle states  in the Born-Oppenheimer approximation, is  
   \begin{equation}
H^{\rm  zero} =   - \frac{1}{2}   \frac{d^2 } {d\theta^2}+ 
 \frac{\xi^2}{4g^2}  [1- \cos (2 g  \theta) ]   
\label{zero3b}
\end{equation} 
The asymptotic form of the ground state energy, $ {\cal E}_0(g^2)$, at large-orders in perturbation theory is 
evaluated in \cite{Stone:1977au}, using methods developed by Bender and Wu  
\cite{Bender:1969si}:
 \begin{equation}
 {\cal E}_0(g^2)  \equiv 
E_0 \xi^{-1}   = \sum_{q=0}^{\infty} a_{0,q} {(g^2)}^{q}, \;\;  a_{0,q} \sim - \frac{2}{\pi} \left( \frac{1} {4\xi}\right)^{q} q! 
\label{bw}
 \end{equation}
The series is ``Gevrey-1'' \cite{Ecalle:1981}, non-alternating, and hence non-Borel summable;  a manifestation of the fact that we are expanding the ground state energy  along a Stokes ray in the complex-$g^2$ plane. The Borel transform is given by $B   {\cal E} (t) = 
- \frac{2}{\pi}  \sum_{q=0}^\infty \left(  \frac{t}{ 4 \xi} \right)^q   
 = - \frac{2}{\pi} \frac{1}{1- \frac{t}{4 \xi}} $, 
and has a pole  singularity   on the positive real axis $\R^{+}$ (i.e., non-Borel summability or ambiguity of the sum).  However, the series is  {\it right}- and {\it left}-  Borel resummable, given by  
$ {\cal S}_{0^\pm }  {\cal E}  (g^2) = \frac{1}{g^2} \int_{C_{\pm}} dt  \; B {\cal E}  (t)  \;  e^{-t/g^2}$, where the contours $C_{\pm}$ pass  above (below) the singularity. Equivalently, taking  $g^2 \rightarrow -g^2$, the series (\ref{bw}) becomes  Borel summable. Analytically continuing the sum  back along $\tilde C_{\pm}$ yields the two ``lateral'' Borel sums:
 \begin{align} 
{\cal S}_{0^\pm }  {\cal E}  (g^2) 
= \Re\,  \B_0   \mp  i  \frac{16 \pi}{  g^2N } e^{-\frac{8 \pi}{g^2N}} 
\label{s1}
\end{align}
where the  imaginary part is the leading non-perturbative ambiguity of resummed perturbation theory, which is $O(e^{-S_ {[\K_{i}  \bar \K_{i} ]}}) \sim e^{-2 S_ {0}} \sim  e^{-2S_I/ N}$,  and  
$ \Re\,  \B_0 \sim O(1) $ is the unambiguous real part. 

We now come to the crux of the matter: the resummed vacuum energy has an imaginary part, but it is {\it not} associated with the Dyson {\it instability}, or {\it decay of the vacuum}.  Rather, this ambiguous imaginary part is a direct reflection of the fact that  (resummed)  perturbation theory by itself is ill-defined. Furthermore, the ambiguity that we find for \cpn on $\R \times \mathbb S^1_L$ is parametrically  the same as that of the elusive IR-renormalons. Therefore, what was viewed as a problem,  in fact becomes a blessing in disguise:
consider the  $\Theta$-independent part of the  vacuum energy density in  a trans-series expansion  (combining perturbative and non-perturbative terms), and collect unambiguous terms and ambiguous terms together: 
 \begin{align}
 \label{conf}
&  {\cal E}_{0, \rm transseries}(g^2) \cr
&= \sum_{q=0}^{\infty} a_{0,q} {(g^2)}^{q} +   \,[\cB_{ii}]   \sum_{q=0}^{\infty} a_{2,q} {(g^2)}^{q}  + \ldots  ({\rm formal})
\cr 
&   \longrightarrow 
\B_{0, \th=0^\pm}+  [\cB_{ii} ]_{\th=0^{\pm}} \B_{2, \th=0^\pm}  + \ldots ({\rm BE-resummation}) \cr  
 &  =  \Re \, \B_0   +   \Re\,[\cB_{ii} ]  \Re \, \B_2  + i [ \Im\,\B_{0, \th=0^\pm}  +    \Re \, \B_2  \Im\, [\cB_{ii}]_{\th=0^{\pm}} ] \qquad \cr
 & =    \Re \, \B_0   +   \Re\,[\cB_{ii} ]  \Re \, \B_2    ,\quad {\rm up \; to}  \; e^{-4S_0}  \qquad  \qquad  \qquad 
 \end{align}
In our explicit computation,  we have taken $ \B_2  = a_{2,0} + O(g^2)$, and kept only  $a_{2,0}$ for consistency because  we are also only accounting for the leading  large orders asymptotics in  (\ref{bw}). 
The sum of the left (right)  Borel resummation of perturbation theory  and non-perturbative left (right) neutral  bion amplitude is {\it unambiguous} at order  $e^{-2S_0}= e^{-2S_I/N}$, as encoded in our perturbative-non-perturbative ``confluence equation" in (\ref{conf}):
\begin{align} 
 \Im\,\B_{0, \th=0^\pm}  +    \Re \, \B_2  \Im\, [\cB_{ii}]_{\th=0^{\pm}} =0, \; {\rm up \; to}  \; e^{-4S_0}  \
 \label{confluence} 
\end{align} 
 The passage  from $\theta=0^{-}$ to   $\theta=0^{+}$ is accompanied by a ``Stokes jump" for the Borel resummation (\ref{s1}), which is mirrored by a jump in the  neutral bion amplitude  in the opposite direction  
 (\ref{bion-ambiguity}) such that the sum of the two  gives a unique result, with a smooth limit up to ambiguities at order $e^{-4S_0}$.   Eq.(\ref{confluence}) is conjectured to hold in (deformed) Yang-Mills in  \cite{Argyres:2012vv,Argyres:2012ka}, and here we verify it by explicit computation for \cpn. 
 Confluence equations are   crucial for giving a non-perturbative continuum definition of QFT.  We refer to the  procedure in (\ref{conf})  as {\it Borel-\'Ecalle resummation}, after \'Ecalle's seminal work \cite{Ecalle:1981}, which formalized asymptotic expansions with exponentially small terms  (trans-series) and generalized Borel resummation to account for  the Stokes phenomenon. 

As an application, we calculate a physically interesting non-perturbative quantity.
The mass gap  is the energy required to excite the system from the ground state to the 
 first excited state. For \cpone,  in the standard notation for Mathieu functions, the pair of states 
 ${\rm ce}_n (\theta, q)$  and    ${\rm se}_{n+1} (\theta, q)$  ($q=\frac{\xi^2}{4g^4}$), $n=0,1,2, \ldots$, become  degenerate to all orders in perturbation theory: their asymptotic expansions are identical. 
 As $g^2  \rightarrow 0$,    the splitting  
${\cal E}(b_{n+1}) - {\cal E}(a_{n}) $  is purely non-perturbative.  The mass gap is defined 
as $m_g= {\cal E}(b_{1}) - {\cal E}(a_{0}) $ and is given by
 \begin{eqnarray}
\label{massgap1}
m_g =  \frac{8 \pi }{g}   \left(1 - \frac{7 g^2 }{16 \pi}  + O(g^4) \right) e^{-\frac{2 \pi}{g^2}}   \sim    e^{-S_I/2}  
\end{eqnarray}
This also justifies the Born-Oppenheimer approximation,  because low-lying states are non-perturbatively split,   whereas their separation  from the higher states is  an order one  gap:
  $m_g \ll  {\cal E}(a_1) - {\cal E}(b_1) \sim \Delta {\cal E}_{\phi}$,
where $ \Delta {\cal E}_{\phi}$ is the gap in the $\phi$-sector in  (\ref{zeroth}). 
For   \cpn,   generalizing the above   discussion,  we find 
$m_g  \sim  \frac{1}{\sqrt \lambda}    e^{-\frac{4 \pi}{\lambda}}   \sim e^{-S_I/N} $, which is a kink-instanton effect (\ref{ki}). 
We are not aware of any previous
microscopic derivation in \cpn of the all-important non-perturbative mass gap   $\sim e^{-S_I/N}$.  The  gap   at small-$L$ may be considered as {\it the  germ of the mass gap} for the theory on $\R^2$.   At large-$N$, this agrees with the mass gap obtained by the master field method \cite{D'Adda:1978kp}. 
 
One is entitled to ask whether this result is really meaningful, since  there is a perturbative  series  multiplying the kink-instanton amplitude, see (\ref{massgap1}),  which is  itself a divergent   
asymptotic  (non-Borel summable) series.  This question has been answered in the mathematics literature.  An important  result by Pham et al, and Delabaere  \cite{ddp}, using \'Ecalle's theory of resurgence \cite{Ecalle:1981},  proves that the semi-classical expansions for the energy levels of the QM double-well and periodic potentials are indeed resurgent functions, resummable to finite, unambiguous, exact results. Our primary contribution here  is that we have  found the conditions under which a non-trivial QFT such as \cpn is connected to QM  without any rapid cross-over or phase transition, i.e., by guaranteeing continuity.  This permits us to derive the germ of all non-perturbative observables in QFT   in the small-$\mathbb S^1$ domain using rigorous QM results.  Introducing the 
  $\Theta$  dependence leads to a 
   `grading' of the resurgent trans-series structure \cite{Dunne:2012ae}.   The QFT results are in both qualitative and quantitative agreement with lattice and large-$N$ results.  We hope that this remarkable connection between QFT and QM may be used to explore other non-perturbative properties of general QFTs, and eventually lead to a fully consistent non-perturbative definition of non-trivial QFTs in the continuum.

We acknowledge support from DOE grants DE-FG02-92ER40716 (GD) and  DE-FG02-12ER41806 (MU).

\end{document}